\documentstyle[sprocl]{article}

\def\be{\begin{equation}}
\def\ee{\end{equation}}
\def\bea{\begin{eqnarray}}
\def\eea{\end{eqnarray}}

\begin{document}

\title{Mutual Exclusion Statistics in the Exactly Solvable Model\\
of the Mott Metal-Insulator Transition
}

\author{
Yasuhiro Hatsugai$^{1}$,
Mahito Kohmoto$^{2}$,
Tohru Koma$^{3}$,
and
Yong-Shi Wu$^{4}$}
\address{
$^1$ Department of Applied Physics, University of Tokyo,
Hongo, Bunkyo-ku, Tokyo 113, JAPAN\\
$^2$ Institute for Solid State Physics, University of Tokyo,
 Roppongi, Minato-ku, Tokyo 106, JAPAN \\
$^3$ Department of Physics, Gakushuin University,
Mejiro, Toshima-ku, Tokyo 171, JAPAN\\
$^4$Department of Physics, University of Utah,
Salt Lake City, Utah 84112, U.S.A.
}


\maketitle\abstracts{
We study statistical characterization of the many-body
states in the exactly solvable model with internal degree of freedom
 in more than one dimension. The model exhibits
the Mott metal-insulator transition.  It is shown that
the ground state  is  described
by that of a generalized ideal gas of particles (called
exclusons) which have mutual exclusion statistics
between different species.
In addition to giving a perspective view of spin-charge separation,
the model
constitutes an explicit example of mutual exclusion statistics
in more than two dimensions.
}

\section{Introduction}
\label{Intro}
\noindent
Elementary particles or excitations are usually classified either as
boson or as fermion. In recent years, however, it has been recognized
that particles with ``fractional statistics" intermediate between boson
and fermion can exist in two-dimensional \cite{Wilczekbook} or in
one-dimensional \cite{YangYang,Haldane} systems. In two dimensions, a
type of fractional statistics can be defined on the basis of the phase
factor, $\exp(i\theta)$ with $\theta$ allowed to be arbitrary,
associated with an exchange of identical particles. One has $\theta=0$
for bosons and $\theta=\pi$ for fermions. A particle obeying such
fractional statistics (with $\theta\not=0$ or $\pi$) is called as
``anyon" \cite{wil}. It is believed that the quasiparticles
and the quasiholes in the fractional quantum Hall liquids are
anyons \cite{Halperin}. Anyons can exist only
in two spatial dimensions due to the braid group structure
associated with them \cite{Braidgroup}.

Another aspect of quantum statistics involves state counting, or the
exclusive nature of the particles. Any number of bosons can be in a
single-particle quantum state. Therefore there is no exclusion between
bosons. On the other hand, the exclusion is perfect for fermions in a
sense that a single particle state can accommodate at most one
fermion. This aspect of quantum statistics can be generalized, as
noticed by Haldane \cite{Haldane}, who proposed a definite
generalization of Pauli principle such that one can consider particles
with non-perfect exclusion. He pointed out that a spinon in
one-dimensional long-range interacting quantum spin chain can exclude,
on average, half of other spinon in occupying a single particle state.
We shall call such a generalization as ``exclusion statistics" and a
particle obeying it as an ``excluson". In contrast to usual bosons and
fermions, the general concept of exclusons allows mutual statistics.
Namely, there may exist statistical interactions or mutual exclusion
between different species of particles. Haldane has recognized
\cite{Haldane} that quasiparticles in the fractional quantum Hall
fluids are exclusons with mutual statistics between quasi-electrons
and quasi-holes. More detailed discussions that show the differences
and relationship between the anyon description and the excluson
description of quasiparticles in the fractional quantum Hall fluids
are given in Ref. 7. This gives a concrete example of an excluson
system in two dimensions.

Thus the concepts of fractional anyon statistics and exclusion
statistics constitute generalizations of two different aspects
(exchange phase and exclusion) of usual quantum statistics. An
essential difference between the two concepts is that anyons can
exist only in two spatial dimensions, while in principle exclusons
may exist in any dimensions.

Recently, one of us\cite{Wu} introduced the concept of generalized
ideal gas of exclusons (see Sec.~\ref{excluson} for definition),
and showed that its thermodynamic properties can be
easily understood through a
statistical distribution that interpolates between bosons and
fermions. Later Bernard and Wu\cite{BernardWu} have shown that the
Bethe ansatz solvable models in one dimension can be described as an
ideal (or non-interacting) excluson gas.  This exemplifies that in
certain circumstances particle-particle interactions can be totally
absorbed by the statistical interactions.
(A possible relation to a conformal field theory was
discussed by Fukui and Kawakami \cite{Kawakami}.)
Thus the concept of exclusion statistics may become a powerful
tool in understanding certain interacting many-body problems.
For example, recently it has been shown \cite{Boson} that the
essential features of low-temperature physics of Luttinger
liquids in one dimension can be approximately described by a
system of noninteracting exclusons. This may provide a
new approach to interacting many-body systems.

The examples considered by Bernard and Wu are the repulsive
$\delta$-function boson gas\cite{delta} and the Calogero-Sutherland
model\cite{1/r2}. Both of them contain only single species. The
present paper studies statistical interactions or
mutual statistics in exactly solvable models with internal quantum
numbers. We shall consider an exactly solvable model in
arbitrary dimensions proposed by two of us\cite{HK}, which has the Mott
metal-insulator transition.  An interesting feature  is
that there exists mutual excluson between different species. The
question we are going to address deals with the excluson
description for the physically interesting phenomenon of charge-spin
separation.  The model under consideration exhibits this phenomenon
under certain conditions.
We shall show that when this happens,  the  model
can be described by two species (spin and charge) of exclusonic
excitations with nontrivial mutual statistics. This provides an
 example of (mutual) exclusion statistics
in more than two dimensions.

\section{Excluson Description}
\label{excluson}
\noindent
We consider a system with a total number $N=\sum_{j,\mu} N_j^\mu$
of particles or quasiparticles,
where $N_j^\mu$ is the number of particles of species $\mu$ with
a set of {\em good} quantum numbers, collectively denoted by $j$,
specifying the states.
Following Ref. 8, we assume that the total number of
states with $\{N_j^\mu \}$ is
\begin{equation}
W=\prod_{i,\mu} {\left[D_i^\mu(\{N_j^\nu\})+N_i^\mu-1\right]! \over
N_i^\mu!\left[D_i^\mu(\{N_j^\nu\})-1\right]!},
\label{Wnumber}
\end{equation}
where $D_i^\mu(\{N_j^\nu\})$ is the number of available single particle
states (counted as bosons), which by definition is given by
\begin{equation}
D_i^\mu(\{N_j^\nu\})+\sum_{j,\nu} g_{ij}^{\mu\nu}N_j^\nu=G_i^\mu,
\label{mutualEq}
\end{equation}
with statistical interactions $g_{ij}^{\mu\nu}$.
Here $G_i^\mu$ is the number of available single particle states
when there is no particle in the system.
Namely, $G_i^\mu=D_i^\mu(\{0\})$. The derivative of
(\ref{mutualEq}) is
\begin{equation}
{\partial D_i^\mu(\{N_j^\nu\}) \over \partial N_j^\nu}
=-g_{ij}^{\mu\nu},
\end{equation}
agreeing with the original definition for ``{\em statistical
interactions}" proposed by Haldane\cite{Haldane}. When the pair
$(i,\mu)$ differs from $(j,\nu)$, we call $g_{ij}^{\mu\nu}$
{\em mutual} statistics between particles labelled by $(i,\mu)$ and
those by $(j,\nu)$.

We further assume that the total energy of the system with
$\{N_j^\mu\}$ particles is always simply given by
\begin{equation}
E=\sum_{j,\mu}{N_j^\mu}\epsilon_j^\mu
\label{totalE}
\end{equation}
with constant $\epsilon_j^\mu$. Equations (\ref{Wnumber}),
(\ref{mutualEq}) and (\ref{totalE}) define the generalized
ideal gas of exclusons \cite{Wu}. It is known that (\ref{totalE})
is not satisfied for free anyons \cite{nosum}.
One of us \cite{Wu} has derived statistical distribution for
generalized ideal gas and the thermodynamics following from it.
The equilibrium statistical distribution for $\{N_{i}^{\mu}\}$
is determined by
\begin{equation}
w_i^\mu N_i^\mu + \sum_{j,\nu}g_{ij}^{\mu\nu}
N_{j}^{\nu} = G_i^\mu,
\label{eq_for_w}
\end{equation}
where $w_i^\mu$ satisfy the  equations
\begin{equation}
(1+w_i^\mu ) \prod _{j,\nu}
{\left({\frac {w_j^\nu}{1+w_j^\nu} } \right)}^{g_{ji}^{\mu\nu}}
= \exp \left[{\frac {\epsilon_i^\mu-a^{\mu}} {T}}\right],
\label{eq_det_w}
\end{equation}
where $a^\mu$ is the chemical potential for particles of
species $\mu$.
The thermodynamic potential is given by
\begin{eqnarray}
\Omega & \equiv & -T \log Z\\
& = & -T \sum_{\mu,i} G_i^\mu \log \left[
{
\frac
{G_i^\mu +N_i^\mu-\sum_{j,\nu} g_{ij}^{\mu\nu} N_j^\nu}
{G_i^\mu -\sum_{j,\nu} g_{ij}^{\mu\nu} N_j^\nu}
}\right],
\label{OmegaExdes}
\end{eqnarray}
where $Z$ is the grand partition function.

\section{Exactly Solvable Model in Higher Dimensions}
\label{HKmodel}
\noindent
 In two dimensions,
it is known that the quasiparticles of the fractional quantum Hall
liquid are anyons. They can also be considered to be exclusons
\cite{qpFQHE}. In this section we present an example of mutual
exclusion between different species in an exactly solvable model
in {\em higher dimensions}, that exhibits charge-spin separation
under certain circumstances. This clearly shows that exclusion
statistics is conceptually different from anyon statistics
whose existence requires two (spatial) dimensions.

Recently two of us \cite{HK} and Baskaran \cite{Baskaran} have
proposed a model of interacting electrons that can be solved
exactly in any dimensions. The Hamiltonian is
\begin{equation}
H= -\sum_{\langle i,j\rangle}c_{i,\sigma}^\dagger c_{j,\sigma} + h. c.
+{\frac U {L^d}}\sum_{i,j,\ell,m}\delta_{i+j,\ell+m}
c_{i,\uparrow}^\dagger c_{j,\uparrow}
c_{\ell,\downarrow}^\dagger c_{m,\downarrow}
- \sum_{i,\sigma} (\mu + \sigma \mu_0 h )
c_{i,\sigma}^\dagger c_{i,\sigma},
\label{HKHamil}
\end{equation}
where $\langle i,j\rangle$ represents nearest neighbors in $d$
dimensions, and $L^d$ is the total number of lattice sites,
$\mu$ the chemical potential, $\mu_0$ the magnetic moment and $h$
the external magnetic field.

This model is unrealistic in the sense that the interaction term
with coefficient $U$ is of infinite-ranged in real space and of
strength independent of distance. It, however, has an attractive
feature of being exactly solvable. In fact, it can be easily
diagonalized for each $k$ in momentum space. All the properties
including the thermodynamic quantities were obtained
in \cite{HK}. Furthermore
it is remarkable that this model exhibits a number of important
features of the correlated electron problems in spite of
its simplicity and unrealistic nature. It exhibits, for example,
the Mott metal-insulator transition that was stressed by two of
us \cite{HK} and  Continentino and Coutinho-Filho \cite{Brazilpaper}.

The zero temperature phase diagram of this model in any dimensions
is shown in Fig.~\ref{fig}. It has both the fixed-density and
density-driven Mott transitions. These transition in general may
be in different universal classes \cite{Continentino,Fisheretal}.
The critical exponents of the two types of transitions are, however,
the same in the present model and they seem to be in the same
universality class \cite{Brazilpaper}. In the zero temperature
phase diagram Fig.1, the region OBC is a Mott insulator phase with
half filled band. The rest is metallic phases. In the region OABC,
a double occupancy is prohibited and, as we will show, the
system can be described by an excluson picture.
There is a Fermi surface of the excluson gas in the region OAB ,
and on the phase transition line OB is a quantum phase transition
of the excluson gas. On the other hand, the excluson description
breaks down on the phase transition line BC. The region outside
OABC is a metallic phase which is described by the two species of
fermions (spin up and spin down electrons) as it should be.

Let us concentrate on the region OAB. Assume that $U$ is large
and $T$ is low, so that $U$ is much larger than both $T$ and
the band width. Under these conditions, there is no activation
of doubly occupied states, therefore there are only
three states (0,1 and 2) for each momentum $k$.
In the state 0 there is no electron, the state 1 an electron
with spin up, and the state 2 an electron with spin down.
Let us denote the number of charges as $N_c$ and the number
of magnons (number of spin-down) as $N_s$. We regard $N_c$
and $N_s$ as independent variables (spin-charge separation).
By definition, the state 0 has $N_c=0$ and $N_s=0$, the state 1
$N_c=1$ and $N_s=0$ and the state 2 $N_c=1$ and $N_s=1$
(See Table~\ref{table1}). Then from (\ref{mutualEq}), we
easily derive
\begin{equation}
G_c=1 \ , \ G_s=0,
\end{equation}
and
\begin{equation}
\left[
\matrix{g_{cc}&g_{cs}\cr
g_{sc}&g_{ss}}
\right] =
\left[
\matrix{1&0\cr-1&1}
\right] \; .
\label{gHK}
\end{equation}

\begin{table}[t]
\caption{Electronic states and spin charge labels}
\label{table1}
\begin{tabular}{cccc}
\hline\hline
Electronic State & Label & {$N_c$: Charge } & {$N_s$: Magnon} \\
\hline
$0$&$0$&$0$&$0$\\
$\uparrow$&$1$&$1$&$0$\\
$\downarrow$&$2$&$1$&$1$\\
\hline\hline
\end{tabular}
\end{table}

It is easy to verify that the condition for the ideal excluson
gas (\ref{totalE}) is satisfied, so that the system with doubly
occupied states suppressed can be described as a generalized ideal
gas with two (charge and magnon) species with the statistical
interaction given by (\ref{gHK}). Note the nontrivial value $-1$
for mutual statistics $g_{sc}$; i.e.\  the presence of a charge
can creat an available state for magnon, though there is no bare
available single magnon state ($G_{s}=0$) when there is no charge.
It is straightforward to check that the thermodynamics of the
generalized ideal excluson gas obtained from Eqs.~(\ref{eq_det_w})
and (\ref{OmegaExdes})
is identical to the result of Ref. 14 in the low
temperature limit.

Indeed, in the present case, the species index $\mu=c,s$, and
the state index $j$ is the momentum $k$ in $d$-dimensional space.
Equation~(\ref{OmegaExdes}) now takes the form
\begin{equation}
\Omega = -T\sum_\mu\int
{\frac {d^dk}{(2\pi)^d}}
\log\left[
\frac
{1-n_\mu(k)-\sum_\nu g_{\mu\nu} n_\nu(k)}
{1-\sum_\nu g_{\mu\nu} n_\nu(k)}\right],
\label{OmegaHK}
\end{equation}
where $n_\mu(k)$ is the occupation number distribution function of
the charge ($\mu=c$) or spin ($\mu=s$) excitations in
$k$-space. From Eqs.~(\ref{eq_for_w}) and (\ref{eq_det_w}), we have
\begin{equation}
n_c(k)[1+w_c(k)] = 1, \ \ n_s(k)[1+w_s(k)] = n_c(k),
\label{HKn}
\end{equation}
where the statistics matrix (\ref{gHK}) has been used , and
$w_c(k)$, $w_s(k)$ satisfy
\begin{eqnarray}
w_c(k) {\frac{1+w_s(k)}{w_s(k)}} &=& e^{(\epsilon_c(k)-\mu_c)/T}, \\
w_s(k) &=& e^{(\epsilon_s(k)-\mu_s)/T},
\label{HKws}
\end{eqnarray}
where $\epsilon_c(k)=-2\sum_{\alpha=1}^d \cos(k_\alpha) $,
$\epsilon_s=2\mu_0h$
(the energy  of spin excitation, which is actually
measured relative to the energy of spin-up electrons), and
\begin{equation}
\mu_c=\mu+\mu_0 h,\;\; \mu_s=0.
\end{equation}
Thus
\begin{equation}
w_c(k)=\frac
 { e^{(\epsilon_c(k)-\mu_c)/T}} {1+e^{-(\epsilon_s(k)-\mu_s)/T}}.
\label{HKwc}
\end{equation}
Substituting Eqs.~(\ref{HKwc}), (\ref{HKws}), and (\ref{HKn}) into
(\ref{OmegaHK}),
we obtain
\begin{eqnarray}
\Omega&=&-T\int {\frac {d^dk}{(2\pi)^d}}
\log\left[1+(e^{\mu_0h/T} + e^{-\mu_0h/T})
e^{(\mu_c-\epsilon_c)/T}\right] \nonumber \\
&=&-T\int {\frac {d^dk}{(2\pi)^d}}
\log\left[1+e^{-(\epsilon_1(k)-\mu)/T}
+ e^{-(\epsilon_2(k)-\mu)/T}\right],
\label{HKresult}
\end{eqnarray}
where $\epsilon_1(k)=\epsilon_{c}(k)-\mu_0h$,
and $\epsilon_2(k)=\epsilon_{c}(k)+\mu_0h$ are
the energy of spin-up and spin-down electrons respectively.
Equation~(\ref{HKresult}) is nothing but the result of
Ref.~\cite{HK} in the low temperature limit.
(Remember that here we consider
the case with large $U$ and low $T$, so that doubly occupied states
are suppressed.)

Here we emphasize that the concept of spin-charge separation
is crucial. The effects that spin and charge excitations
are not actually independent of each other have been taken care of
by the statistical interaction or mutual statistics between them
in the present formulation. It seems to us that similar situations
may happen in other strongly correlated systems that exhibit
charge-spin separation in higher dimensions.

\section*{Acknowledgments}
This work was supported by Grant-in-Aid,
 from the Ministry of Education
, Science and Culture of Japan.
The work of Y.S. Wu was also supported in part
by U.S. NSF grant PHY-9309458.
 He is also grateful for
 warm hospitality of the Institute for Solid State
Physics, University of Tokyo during his visits.


\section*{References}
\begin{thebibliography}{99}
\bibitem{Wilczekbook} See, {\em e.g., Fractional Statistics and Anyon
Superconductivity}, edited by F.~Wilczek (World Scientific, Singapore,
1989) and references therein.

\bibitem{YangYang} C.~N.~Yang and C.~P.~Yang, J. Math. Phys.
{\bf 10}, 1115 (1969).

\bibitem{Haldane} F.~D.~M.~Haldane, Phys. Rev. Lett. {\bf 66}, 937
(1991).

\bibitem{wil} F.~Wilczeck, Phys. Rev. Lett. {\bf 48}, 1144 (1982).

\bibitem{Halperin} B.~ I.~ Halperin,  Phys. Rev. Lett. {\bf 52}, 1583
(1984).

\bibitem{Braidgroup} Y.~S.~Wu, Phys. Rev. Lett. {\bf 52}, 2103 (1984).

\bibitem{qpFQHE} Y.~S.~Wu, Y.~Yu, Y.~Hatsugai, and M.~Kohmoto
(to be published).

\bibitem{Wu} Y.~S.~Wu, Phys. Rev. Lett. {\bf 73}, 922 (1994).

\bibitem{BernardWu} D.~Bernard and Y.~S.~Wu, in Proc. of Sixth
Nankai workshop (Tianjin, China; June 1994),
ed. by Mo-lin Ge and Y.S. Wu (World Scientific, 1995).

\bibitem{Kawakami} T. Fukui and N. Kawakami, Phys. Rev. B {\bf 51},5239
 (1995).

\bibitem{Boson} Y.S. Wu and Yue Yu, Phys. Rev. Lett.
{\bf 75} (1995) 890.

\bibitem{delta} D.~C.~Mattis and E.~H.~ Lieb, J. Math. Phys. {\bf 6},
304 (1965).

\bibitem{1/r2} F.~Calogero, J. Math. Phys. {\bf 10}, 2191, 2197 (1969);
B.~Sutherland, J. Math. Phys. {\bf 12}, 251 (1971).

\bibitem{HK} Y.~Hatsugai and M.~Kohmoto, Physica C {\bf 185-189},
 1539(1991); J. Phys. Soc. Jpn. {\bf 61},
2056 (1992).

\bibitem{nosum}Y.~S.~Wu, Phys. Rev. Lett. {\bf 53}, 111 (1984).


\bibitem{Baskaran} G.~Baskaran, Int. J. Mod. Phys. Lett.
B {\bf 5} 643, (1991).

\bibitem{Brazilpaper} M.~A.~Continentino and M.~D.~Coutinho-Filho,
Solid State Commun. {\bf 90}, 619 (1994).

\bibitem{Continentino} M.~A.~Continentino, Physics Report,
{\bf 239}, 179 (1994).

\bibitem{Fisheretal} M.~P.~A. Fisher, P.~B. Weichman, G.~Grinstein,
and D.~Fisher, Phys. Rev. B {\bf 40}, 546 (1989).
\end{thebibliography}
\end{document}